\documentclass[aip,amsmath,amssymb,reprint]{revtex4-1}

\usepackage[dvipsnames]{xcolor}
\usepackage{graphicx}
\usepackage{dcolumn}
\usepackage{bm}
\usepackage{dcolumn}
\usepackage{epstopdf}
\usepackage{xcolor}
\usepackage{natbib}
\usepackage{subcaption} 
\usepackage{float}
\usepackage{tabularx}
\usepackage{comment}
\usepackage{mathrsfs}
\usepackage{upgreek}
\usepackage[a]{esvect}
\setcitestyle{square}
\usepackage[toc,page]{appendix}
\usepackage[font={small,normalfont}]{caption}
\usepackage[colorlinks=true, allcolors=blue]{hyperref}
\usepackage[normalem]{ulem}
\usepackage{upgreek}
\usepackage[capitalise]{cleveref}
\usepackage{amssymb}
\usepackage{algorithm}
\usepackage{algpseudocode}
\usepackage{natbib} 
\setcitestyle{numbers}
\usepackage{multirow}

\usepackage{makecell}

\begin{document}


\title{Rapid assimilation of high-Z impurity ions along the magnetic field line from an ablated pellet}


\author{Haotian Mao}
\affiliation{Department of Mechanical and Aerospace Engineering, University of California at San Diego, La Jolla, California 92093, USA}
\affiliation{Theoretical Division, Los Alamos National Laboratory, Los Alamos, New Mexico 87545, USA}
\author{Yanzeng Zhang}
\author{Xianzhu Tang}
\affiliation{Theoretical Division, Los Alamos National Laboratory, Los Alamos, New Mexico 87545, USA}


\begin{abstract}
The assimilation of ablated high-Z impurities into the hot surrounding
plasma along the magnetic field is investigated by first-principles
kinetic simulations. It is found that the assimilated impurity ions,
primarily driven by the ambipolar electric force, propagate steadily
into the surrounding plasmas. The high-Z impurities in different
charge states are mostly aligned due to the strong collisional
friction among them so that the averaged impurity ions charge
$\bar{Z}$ is a deciding factor. Such assimilation is led by an
impurity front that is behind the cooling front due to a smaller
charge-mass-ratio of the impurity ions $\bar{Z}/m_I$.  With the help
of a self-similar solution, the speed of the impurity front $U_s$ is
shown to be primarily set by the hot surrounding plasma temperature
$T_0$ with a weak dependence on the pellet plasma temperature,
underscoring the collisionless nature of the impurity assimilation
process. Specifically, $U_s\sim \sqrt{\bar{Z}T_0/m_I}$. The
ambipolar-constrained electron conduction flux from the hot plasma is
primarily responsible for the collisionless impurity assimilation
process.
\end{abstract}

\maketitle

\section{\label{sec:section1}Introduction}

High-Z pellet injection has become the method of choice in disruption
mitigation of thermal quench (TQ) in tokamak reactors
\cite{Taylor_disruption_1999,shiraki_2016,
  Gebhart_2021,baylor2019shattered,lehnen2015disruptions,sweeney2020mhd}.
The idea is to have high-Z impurities radiate away the plasma thermal
energy, mostly through line emissions, so that the plasma power flux
reaching the divertor and first wall is
minimized~\cite{Meitner_2017,Sugihara-etal-nf-2007,Hollmann_2014,commaux2016first}. There
are two essential issues related to pellet injection that must be
addressed for reactor applications in TQ mitigation, namely: (i)~how
deeply the pellets can penetrate into a magnetized plasma at fusion
conditions; and (ii)~how uniformly the high-Z impurities can be
assimilated into the fusion plasma along the toroidal direction, which
sets the radiation asymmetry around the
torus~\cite{Lehnen-etal-nf-2015}. The resulting toroidal peaking
factor measures the localization of the radiative power load on
segments of the first wall, and must be controlled within certain
limit to ensure the integrity of the first wall in a mitigated thermal
quench.

Much work has gone into (i)~in the form of pellet ablation models,
which translate into a pellet mass deposition profile after
accounting for the pellet passage through the plasma at a given pellet
velocity.  This ablation rate is driven by the plasma energy flux
deposited into and onto the pellet. Interesting phenomena include
vapor shielding, which can drastically reduce the power flux from the
plasma that can reach the solid
surface~\cite{Parks_1977,PhysRevLett.94.125002,
  Houlberg_1988,Pegourie_2005}. Through ablation, a passing pellet
would leave behind a trail of ablated pellet gas cloud, which gives rise to
the issue (ii)~of impurity ion assimilation (mixing) into the
surrounding plasma. This is the topic of current paper.
Impurity assimilation begins with an isotropic expansion of
the ablated pellet gas cloud as neutral particles do not interact with
electromagnetic fields. This expansion is quickly stopped across the
magnetic field as the impurity gas cloud is ionized by surrounding plasmas, and
further assimilation of the impurity ions is primarily along the
magnetic field which is in the toroidal direction.

\begin{figure*}
    \centering
    \begin{minipage}[b]{\textwidth}
    \includegraphics[width=0.9\textwidth]{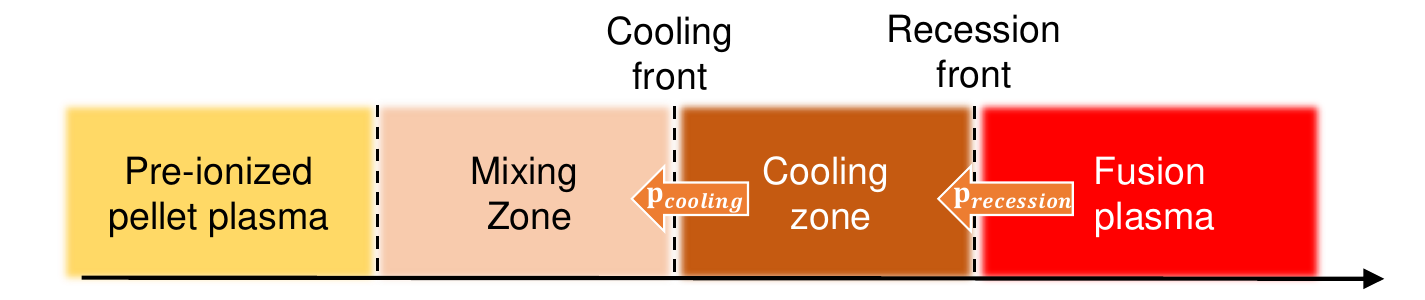}
    \end{minipage}
    \caption{Schematic picture of the impurity ion assimilation process. The orange arrow denotes the energy fluxes that cross the fronts. }
    \label{fig:fig1}
\end{figure*}

The normal expectation is that the parallel transport of impurity
ions is driven by the plasma power flux from the surrounding
high-temperature but low-density plasmas into the much colder and
denser pellet ions.
Since previous studies have mostly used fluid or MHD models, the
plasma power flux is usually modeled by Braginskii closures, with
possibly flux limiting in the initial high-temperature phase to
approximate the long-mean-free-path effect on heat
flux~\cite{Jorek_2014,nimrod2019}. However, recent work reveals the
subtle parallel transport physics that can regulate the plasma power
flux reaching the impurity ions and thus affect the surrounding
plasma cooling~\cite{Zhang_2023,zhang2023electron}, the essential
complication of which is indicated by various fronts in
Fig.~\ref{fig:fig1} (notice that an artificial boundary that mimics
the radiative pellet cloud was previously used in
Ref.~\cite{Zhang_2023,zhang2023electron}).

Specifically, a rarefaction wave forms, producing a recession front
that moves away from the ablated pellet mass and into the surrounding fusion
plasma. Behind the recession front, the surrounding plasma will be
accelerated to form a cooling flow toward the ablated pellet mass. The cooling flow
goes through a shock at the cooling front, after which the ion flow
energy is converted into ion thermal energy. One of the most striking
findings is that the plasma power flux through the cooling zone, which
is between the cooling front and recession front, is regulated by
ambipolar transport: the ion power flux is carried by the cooling flow
as convective flow energy and convective thermal energy flux; and the
electron power flux is dominated by parallel thermal conduction, which
is carried by the asymmetry in the distribution function due to the
electrostatic trapping and infalling cold electrons via
ambipolar electric field in the cooling zone. In the early phase, with
a time duration set by the thermal electron toroidal transit time, the
resulting parallel electron thermal conduction flux has a
flux-limiting form proportional to the electron thermal speed. In
latter period, especially after the toroidal transit period of the
recession front, the electron conduction flux would be greatly reduced
to have a convective scaling with the ion parallel
flow~\cite{Li_2023}.

The primary finding of current paper is that the
ambipolar-transport-constrained plasma power flux ($P_{cooling}$ in
Fig.~\ref{fig:fig1}) plays a critical role in the impurity
assimilation process along the magnetic fields.  Moreover,
$P_{cooling}$ is dominated by the escaping hot electrons from the
surrounding plasma into the pellet plasma.  The electron heat flux and
consequently the electron heating behind the cooling front are
essentially collisionless processes that are regulated by ambipolar
transport.  As the result of the collisionless {\em mix and heating}
of electrons behind the cooling front, the impurity assimilation is
driven by an ambipolar electric field, which accelerates the impurity
ions to form an impurity front that is behind the cooling front.  As
the result, the impurity propagation, radiation, and further
ionization would all occur in a ``mixing zone'', where impurities mix
with the surrounding plasma ions, behind the cooling front (e.g., see
Fig.~\ref{fig:fig1}). In this regard, the current work also provides a
validation of the propagating front
physics~\cite{Zhang_2023,zhang2023electron}, from self-consistent
simulations without introducing simplified radiative boundary
condition that was previously deployed.

Specifically, our first-principles kinetic simulation studies with
VPIC~~\cite{5222734,K_J_Bowers_2008,K_J_Bowers_2009} focus on impurity
assimilation along the magnetic field, so the problem setup has a
dense and cold impurity ion cloud that is surrounded by hot fusion
plasmas, as shown in Fig.~\ref{fig:fig2}(a).  The simulation studies
reveal that the impurities front speed is primarily set by the hot
surrounding plasma temperature as opposed to the much colder pellet
plasma temperature.  The impurity front stays behind the cooling front
due to a smaller charge-mass-ratio of the impurity ions than that of
the Deuterium in a fusion plasma.

The rest of the paper is organized as follows: In
Section~\ref{sec:section2}, we introduce the setup of the simulations,
the result of which will be shown in Section~\ref{sec:section3}, and
the physics is elucidated with the help of a self-similar
solution. The electron temperature that determines the impurity front
speed will be discussed in
Section~\ref{sec:Te}. Section~\ref{sec:summary} will conclude.

\section{\label{sec:section2}1D3V VPIC simulation setup}

This section presents the setup of the 1D3V kinetic simulations
with the VPIC code. As aforementioned, the focus of the paper is the assimilation of
high-Z impurities, which are assumed to be pre-ionized $I^{Z+}$ with a
large charge state $Z\gg 1$, into the hot surrounding
plasma, along the magnetic field lines. The key length parameters in such process would be the system
length $L$, ablated pellet cloud size $L_p$, and plasma
collisional mean-free-path $\lambda_{\text{mfp}}$. These length
scales can be contrasted with the plasma Debye length $\lambda_{De}$
that would need to be resolved in the fully kinetic simulations.  While
$L\sim 2\pi R\sim 10$~m and $L_p\sim 10^{-1}$~m are mostly set by the
reactor geometry and the cross section of the pellet shards,
$\lambda_{\text{mfp}}$ and $\lambda_{De}$ strongly depend on the
plasma temperature and density. Specifically, for the surrounding
fusion-grade plasma of $T_{e}(\text{fusion})\sim 10^4$~eV and
$n_e(\text{fusion})\sim 10^{20}$~m$^{-3}$, $\lambda_{\text{mfp}}
(\text{fusion})\sim 10$~km, and $\lambda_{De}(\text{fusion})\sim
10^{-4}$~m.

The pellet ion cloud, which provides the initial condition for the
kinetic simulation studies of impurity assimilation into the
surrounding plasma along the magnetic field, has uncertainties in the
absence of a detailed simulation study of the pellet ablation and
the subsequent ionization process.  We will start with the
approximation that the pellet ion blob is initially in pressure
balance with the surrounding plasma. The uncertainty for the pellet
plasma temperature $T_e(\text{pellet})$ will be addressed by a parametric scan for the pellet plasma.  As an illustrative
example, for $T_e(\text{pellet})\sim 10$~eV, we will have
$n_e(\text{pellet})\sim 10^{23}$~m$^{-3}$, and $\lambda_{\text{mfp}}
(\text{pellet})\sim 10^{-5}$~m~$\ll L$,
$\lambda_{De}(\text{pellet})\sim 10^{-7}$~m. This reveals that
$\lambda_\text{mfp} (\text{pellet})\ll L_p\ll L\ll
\lambda_\text{mfp}(\text{fusion})$, {\it i.e.,} the hot fusion-grade
plasma is nearly collisionless, while the cold pellet plasma is
collisional. Another key point is that the hot electron mean-free-path
in the cold pellet plasma satisfies $\lambda_{\text{mfp}}^{eh-ec}\sim
\lambda_{\text{mfp}}(\text{fusion})n_e(\text{fusion})/n_e(\text{pellet})\sim 10$~m$
\gg L_p,$ so that the pellet is transparent to the hot
electrons~\cite{aleynikov_breizman_helander_turkin_2019,aleynikov2020energy}. In
contrast, the hot ion slowing down mean-free-path on the cold pellet
electrons is much shorter~\cite{tang2014reduced}
$\lambda_{\text{mfp}}^{ih-ec}\sim
\lambda_{\text{mfp}}^{eh-ec}[T_e(\text{pellet})/T_e(\text{fusion})]^{3/2}\sqrt{m_i/m_e}\sim
10^{-2}$~m. So the cold pellet plasma is not transparent for the
surrounding hot ions~\cite{aleynikov2020energy}. Therefore, the hot
ions can be the main energy source for the {\em collisional} pellet
heating, which is further strengthened by the fact that the
surrounding electrons will transfer energy to the ions through the
ambipolar potential in the recession layer~\cite{Zhang_2023}.

\begin{figure}
    \centering
    \begin{minipage}[b]{0.45\textwidth}
    \includegraphics[width=1\textwidth]{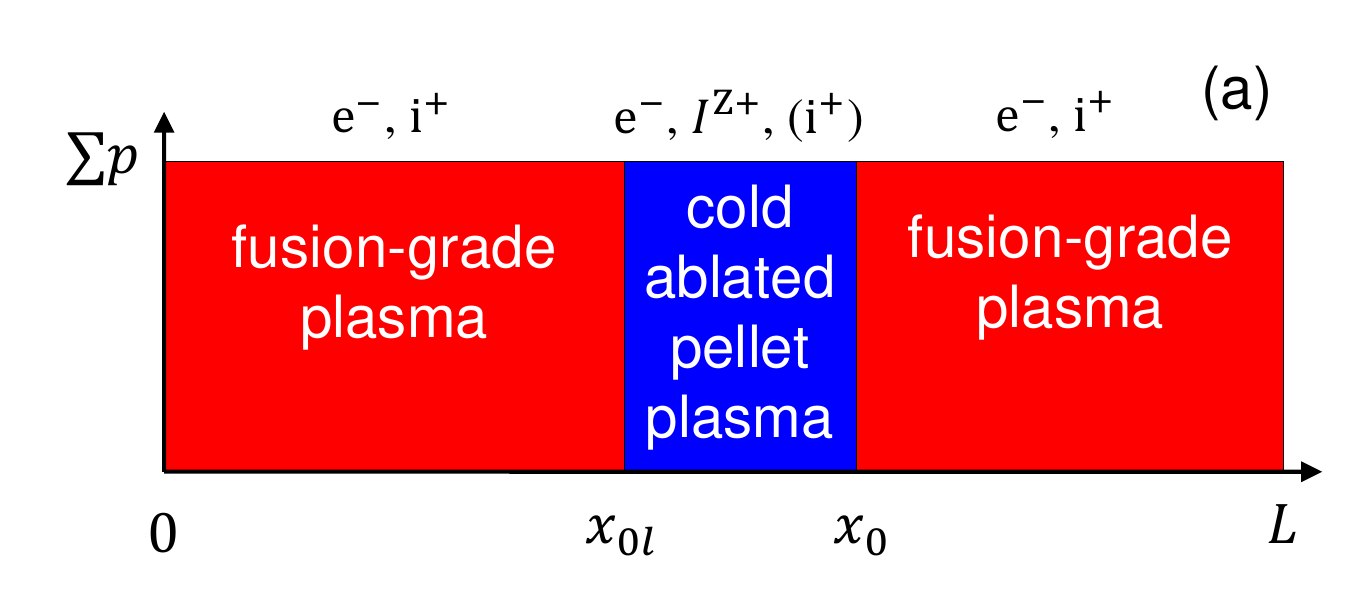}
    \end{minipage}
    \begin{minipage}[b]{0.45\textwidth}
    \includegraphics[width=1\textwidth]{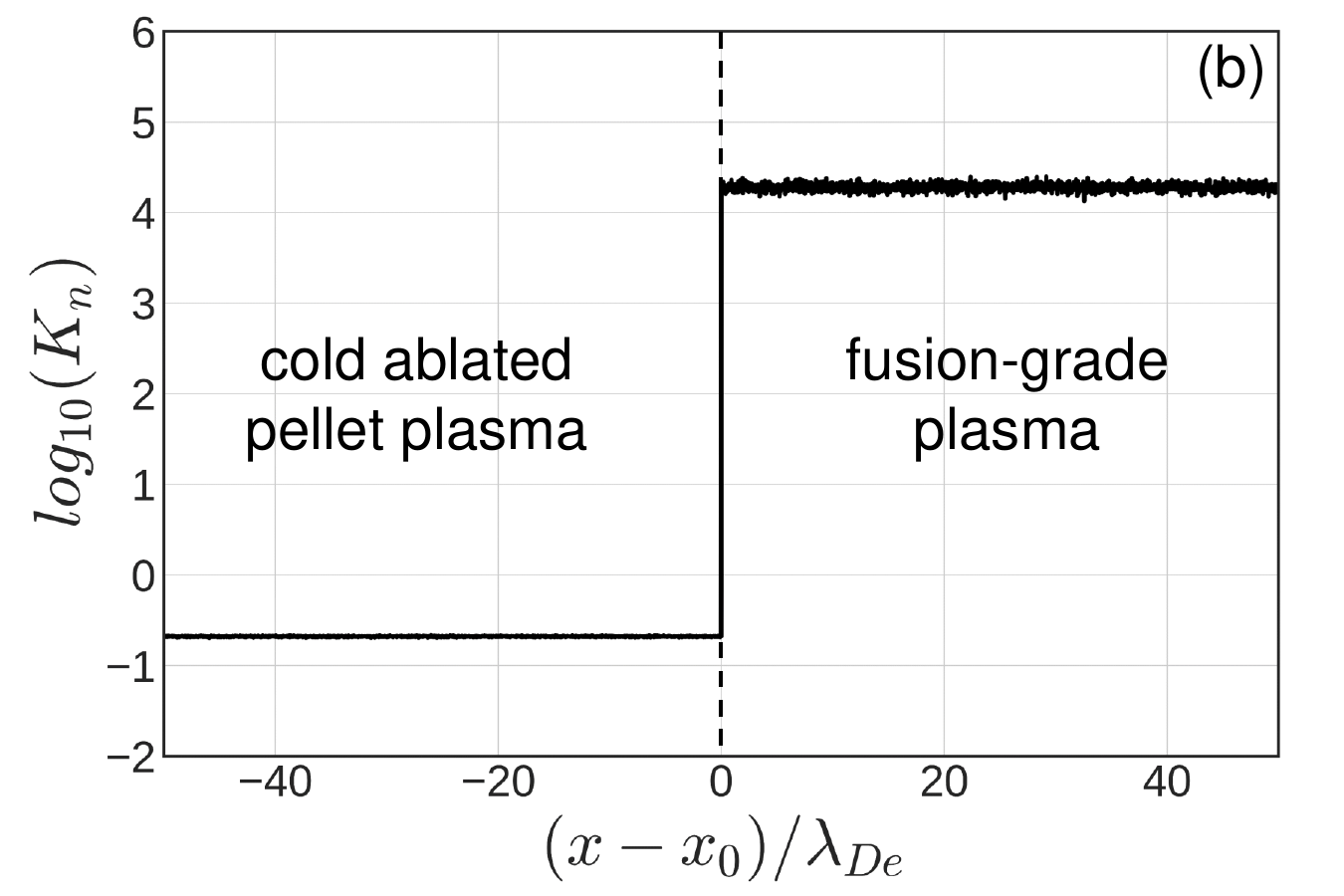}
    \end{minipage}
    \caption{The schematic view of 1D3V kinetic simulation setup~(a) and
      the local Knudsen number $K_n =\lambda_{\text{mfp}}/\lambda_{De}$ in
      logarithm at $t\cdot \omega_{pe} = 0$ for
      $T_{cold}=0.025~T_0$ and single charge state impurities with $Z=7$~(b). Here $\lambda_{\text{mfp}}$ is computed from the local
      electron temperature and density, $\lambda_{De}$ and
      $\omega_{pe}$ are, respectively, the surrounding hot plasma Debye
      length and frequency. }
    \label{fig:fig2}
\end{figure}

A physically meaningful down-scaled kinetic simulation must retain the
aforementioned features, which include (1)~collisionless surrounding
hot plasma; (2)~collisional pellet plasma; (3)~escaping hot electrons
from the surrounding plasma being mostly collisionless in the pellet
plasma; and (4)~escaping hot ions from the surrounding plasma being
collisional in the pellet plasma. A schematic view of such a
simulation setup is given in Fig.~\ref{fig:fig2}. Specifically, we
choose $L=1000~\lambda_{De}$, $L_p=100~\lambda_{De}$ with
$x_{0l}=450~\lambda_{De}$ and $x_{0}=550~\lambda_{De}$ (hereafter, we
will use hot surrounding plasma parameters for normalization and
remove the notation of fusion for simplification). Here a periodic
boundary condition is adopted. In this paper, we focus on an early
stage of the impurity assimilation before the cooling front (and
thus the impurity front as well) reaching the boundary $t\ll
L/c_s^{i}$ with $c_s^i$ the deuterium ion sound speed~\cite{Zhang_2023}.

The simulation studies fix a hot surrounding plasma with temperature
$T_0\sim 10$~keV and density $n_0\sim 10^{20}$~m$^{-3}$, but vary the
cold plasma pellet temperature $T_{cold}\in
[0.01,0.025,0.05,0.1]T_0$. A strong magnetic field along $x$-axis has
been employed in the simulations so that the hot surrounding plasma
$\beta$ is $4\%$. Moreover, an artificial Coulomb Logarithm $\ln
\Lambda$ has been employed in the Coulomb collisions so that the
fusion-grade plasma is collisionless
$\lambda_{\text{mfp}}(\text{fusion})\sim 10^4~\lambda_{De}\gg L$, the
pellet plasma is collisional $\lambda_{\text{mfp}}(\text{pellet})\in
[0.01,0.16,1.25,10]\lambda_{De}\ll L_p$, and the pellet is transparent
for hot electrons $\lambda_{\text{mfp}}^{eh-ec}\in
[100,250,500,1000]\lambda_{De}\gtrsim L_p$.  In our simulations with a
reduced ion mass $m_i=100~m_e$, the hot ion slowing-down
mean-free-path on the cold electrons are
$\lambda_{\text{mfp}}^{ih-ec}\in
[1,10,56,316]\lambda_{De}$. For hot ions, the pellet will be
non-transparent for colder pellet $T_{cold}\leq 0.05~T_0$ but nearly
transparent for hotter pellet $T_{cold}=0.1~T_0$. Interestingly, as we
will show in the next section, the transparentness of the hot ions
will only slightly affect the impurity front speed since the heating
of the pellet electrons is dominated by the collisionless hot
electrons rather than the collisional hot ions.


In the simulation setup, for initial condition we set the total pressure $\sum_j
p_j=\sum_jn_jT_j$ balanced between the hot surrounding plasma and the
cold pellet plasma, where $n_j$ and $T_j$ are the particle
density and temperature, respectively, of species $j$ with a
Maxwellian distribution. The cold pellet plasma is
assumed to be quasi-neutral consisting of a majority of high-Z impurity
ions $I^{Z+}$ and a minority of Deuterium ions $i^+$ with $n_I: n_i = 3:1$
(we will use the scripts $I$ and $i$ to distinguish the impurity and
Deuterium ions). For generality, we consider the cases with
\begin{align}
    m_I/m_i \geq Z,
\end{align}
where $m_I$ and $m_i$ denote the impurity and deuterium masses. For fully ionized high-Z impurities, one should have $m_I/m_i\approx Z$. To make the simulation more efficient, a reduced ion mass of $m_i=100~m_e$ and impurity mass of $m_I=16~m_i$ will be employed. Different impurity charge state $Z$ will be considered. 

It is worth noting that the plasma Debye length is much smaller in the
cold pellet plasma, which will increase in the simulations due to the
pellet plasma heating. Therefore, in the VPIC simulations, we resolve
the cold pellet plasma Debye length $\lambda_{De}(\text{pellet})$,
which ensures the resolution of the cold pellet plasma frequency as
well. Besides, we used 1300 particles per cell for hot plasma, and
thus the cold pellet electrons has $\sim 1300\times T_0/T_{cold}$
particles per cell. Since the impurity density will drop in the mixing
zone, we will cut off the impurity diagnostics at 10 particles per
cell to increase the accuracy.


\section{\label{sec:section3}The assimilation of high-Z impurities along magnetic field line}

The simulations show that the impurities and thus impurity radiative
cooling stays behind the cooling front, in support of the propagating fronts
physics~\cite{Zhang_2023,zhang2023electron} previously found using a
radiative temperature clamping boundary condition away from the
cooling front.  Since the assimilation of the impurities into the
surrounding plasma is inside the mixing zone shown in
Fig.~\ref{fig:fig1}, the electron heat flux there, which is regulated by ambipolar
transport with hot electrons escaping the cooling front mostly
collisionless inside the mixing zone, is expected to play a critical
role in the impurity assimilation process along the magnetic field.

\subsection{Impurity acceleration towards the surrounding hot plasma}

For a cold pellet, the expansion of ions along the magnetic field is driven by the ambipolar electric field~\cite{aleynikov_breizman_helander_turkin_2019,Arnold_2021}, for which the ion charge-mass-ratio $Z/m$ is a key factor. Since the deuterium usually has a larger $Z/m$ than the impurities during the early stage of the expansion where the electron temperature and hence the ionization level of high-Z impurity is low, the deuterium ions will be accelerated faster than the impurity ions without considering their friction force. In fact, it is the competition of the electric force and ion friction force that determines the impurity expansion, which can be illustrated by considering the impurity momentum equation along the
magnetic field line
\begin{align}
    &n_Im_I\frac{dV_{I\parallel}}{dt} =-\frac{\partial p_{I\parallel}}{\partial x} +Z|e|n_IE_x+\sum_{e,i}R_{Ie,i},\label{eq:impurity_momentum_full}
\end{align} 
where $d/dt \equiv \partial/\partial t +
V_{I\parallel}\partial/\partial x$, $n_{I}$ and $V_{I\parallel}$ are,
respectively, the impurity density and its parallel flow velocity,
$p_{I\parallel} = n_{I}T_{I\parallel}$ denotes the impurity ion
pressure with $T_{I\parallel}$ the parallel impurity ion temperature,
and $R_{Ie,i}$ denotes the friction force acted by electrons and
deuterium ions on the impurities.  Notice that the electric force
builds up because the heating of much higher density pellet electrons
will push them into the surrounding plasma. As we will show later, in
the impurity dominant region where the impurity acceleration mainly
occurs, the friction between the electrons and impurities can be
negligible since their momenta are aligned due to the ambipolar
transport constraint. The same equation can be obtained for the
deuterium ions. Notice that the impurity ion pressure provides a weak
drive for the impurity acceleartion compared to the electric
force. This is because on the impurity propagation timescale, the
electron momentum equation satisfies the force balance of
\begin{align}
  -\partial p_{e\parallel}/\partial x-
  n_e|e|E_x +R_{e,I}=0,
\end{align}
with $p_{e\parallel}=n_eT_{e\parallel}$
the electron pressure and $R_{e,I}=-R_{I,e}$. If we assume $p_{I,e}$ develops the same length scale $L_p$ for a
self-similar solution that will be shown in Section~\ref{sec:self-similar}, we have 
\begin{align}
     &\frac{\partial p_{I\parallel}}{\partial x}\sim \frac{n_IT_{I\parallel}}{L_p},\label{eq:efield_scaling}\\
     &Z|e|n_IE_x  \sim \frac{Zn_I}{n_e}\frac{\partial p_{e\parallel}}{\partial x}\sim Z\frac{n_IT_{e\parallel}}{L_p}. \nonumber
\end{align}
Equation~(\ref{eq:efield_scaling}) indicates that the electric
force will be dominant for the impurity acceleration since: (1)~the
electrons in the plasma region get more heating than the impurities
due to the electron conduction heat flux; and (2)~$Z\gg 1$.

From Eq.~(\ref{eq:impurity_momentum_full}), we see that for the pellet with major deuterium and minor high-Z impurity, the impurity acceleration is mainly due to the friction force by the deuterium~\cite{aleynikov2023thermal}. However, for a pellet with major high-Z impurity and minor deuterium, the main drive of impurity ion acceleration will be the electric force, which will be considered in this paper. 

\begin{figure}
	\centering
	\begin{minipage}[b]{0.45\textwidth}
		\includegraphics[width=1\textwidth]{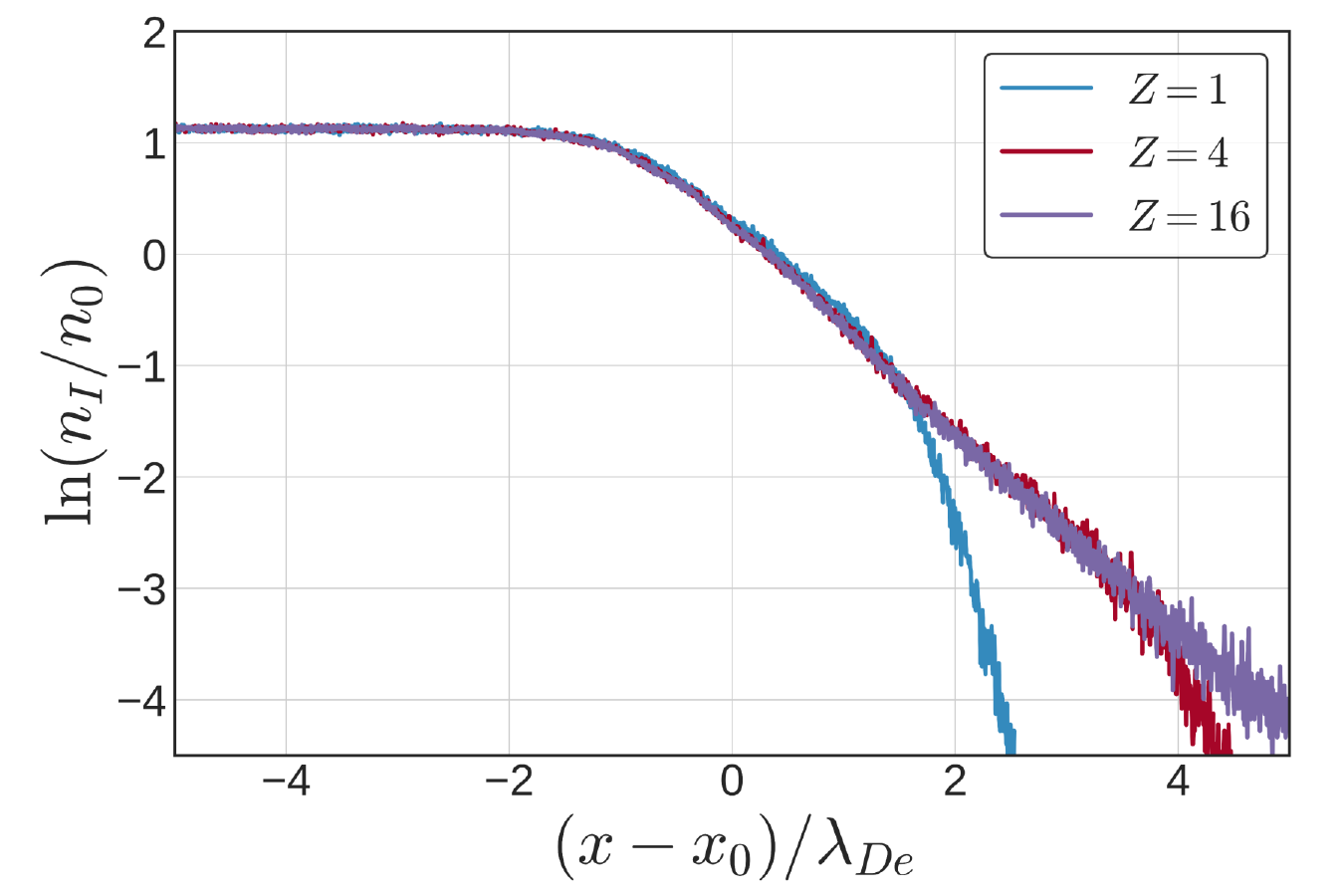}
	\end{minipage}
	\caption{The impurity density of different charge state $Z$ at $t\cdot \omega_{pe} = 101$ for $T_{cold}=0.025~T_0$. Initially the impurity density of different charge state are the same in the pellet.}
	\label{fig:fig2-new}
\end{figure}

Another related consideration would be a multi-components of
impurities, i.e., impurities with different charge state $Z$ due to
varying degrees of ionization. Since the pellet plasma is collisional, the
friction force among the impurities with different charge states
(denoted by $s$) would be large if their density are not negligibly
small since $R_{ss'}\propto n_sn_{s'}Z_s^2Z_{s'}^2$. Such large
friction force would try to align them together, which can be seen
from their momentum equations
\begin{equation}
     \begin{split}
          n_sm_I\frac{d V_{s\parallel}}{d t} =- \frac{\partial p_{s\parallel }}{\partial x} + n_sZ_s|e|E_x + \sum_{e,i}R_{se,i}+ \sum_{s\neq s^{\prime}}{R}_{ss^{\prime}},
     \end{split}
     \label{eq:diff_charge_momentum}    
\end{equation}
where $R_{ss'}\gg n_sZ_s|e|E_x$ for impurities with small $Z_s$. Such
alignment of different impurity components can be seen from
Fig.~\ref{fig:fig2-new}, where we consider a pellet plasma as a mixture
of different charge states $Z=1,4$ and $16,$ that have the same initial
density (so the averaged charge is $\bar{Z}=7$). Notice that when the
impurities density is low enough, the friction force is subdominant,
which is the reason why $Z=1$ will deviate at the very front of the
expansion.

The alignment of the impurity expansion of different charge states
indicates that these impurities share the same flow velocity
$V_{s\parallel}\approx V_{I\parallel}$. As a result, by summing up
Eq.~(\ref{eq:diff_charge_momentum}) of all impurities, the overall
impurity momentum equation can be expressed by the
Eq.~(\ref{eq:impurity_momentum_full}) with $Z$ being replaced by the
averaged charge state $\bar{Z}=\sum_sn_sZ_s/n_I $ and
$n_I=\sum_sn_s$. As such, we will use single charge state impurities
with $Z=7$ for the rest of the simulations and analyses.

It is of interest to note that there are a variety of thermodynamic
forces that can drive the separation of impurities with different
charge states, for example, the ion pressure gradient driven
baro-diffusion, electric field driven electro-diffusion, and electron
and ion temperature gradient driven
thermo-diffusion.~\cite{kagan-tang-pop-2012,kagan-tang-pla-2014,kagan-tang-pop-2014}
But these are all diffusive processes, which are subdominant to the
impurity front propagation into the surrounding plasma that is mostly
collisionless dynamics. This underlies the subtler reason why a
leading order description of the impurity front penetration is given
by the averaged charge of the impurities.  From this angle of
collisional versus collisionless transport, the assimilation of high-Z
pellet ions into the surrounding fusion plasma follows distinctly different
physics as compared with the upstream migration of wall impurities in the scape-off layer.

\subsection{Impurity front}

\begin{figure}
    \centering
    \begin{minipage}[b]{0.45\textwidth}
    \includegraphics[width=1\textwidth]{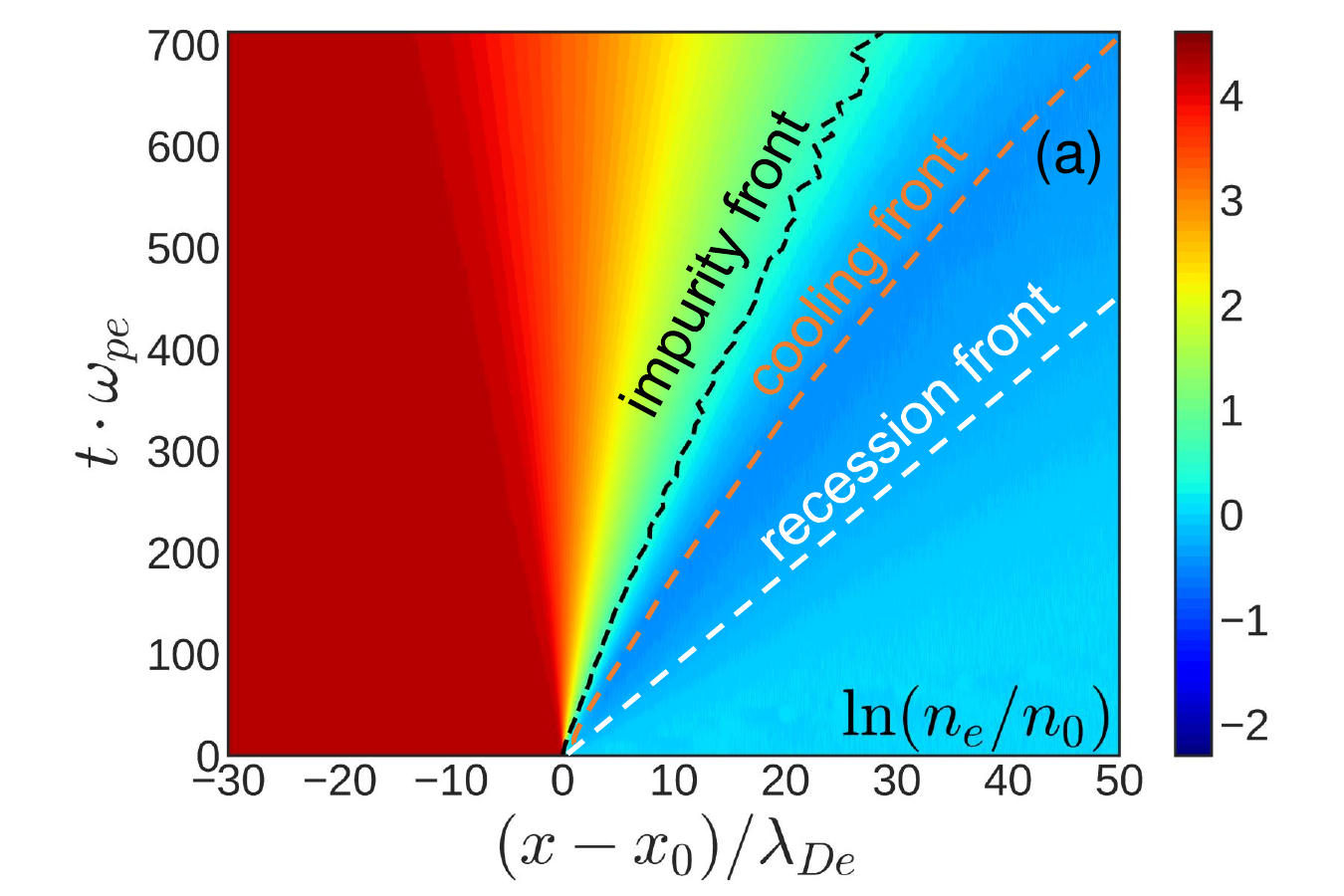}
    \end{minipage}
    \begin{minipage}[b]{0.45\textwidth}
    \includegraphics[width=1\textwidth]{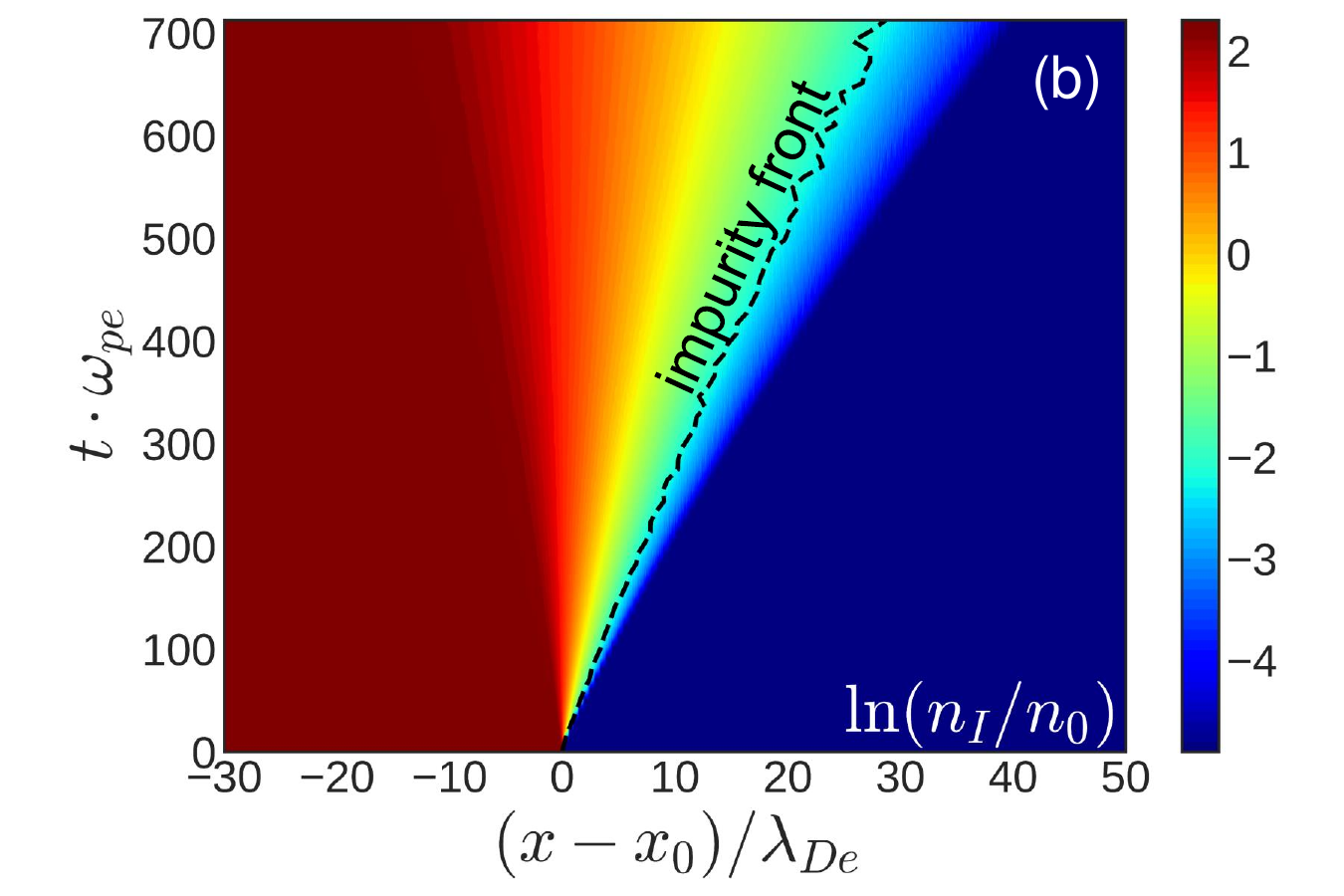}
    \end{minipage}
     \caption{The spatio-temporal evolution of electron density $n_e$~(a) and impurity ion density $n_I$~(b) in logarithm for $T_{cold}=0.025~T_0$ and $Z=7$. The black dash line denotes the impurity front, which is defined in Eq.~(\ref{eq:impurity_front_def}). The orange dash line in the electron density denotes the cooling front (a local minimum in $n_e$) and the white dash line is the recession front where $n_e$ starts to collapse~\cite{Zhang_2023}. }
    \label{fig:fig3}
\end{figure}

\begin{figure}
    \centering
    \includegraphics[width=0.45\textwidth]{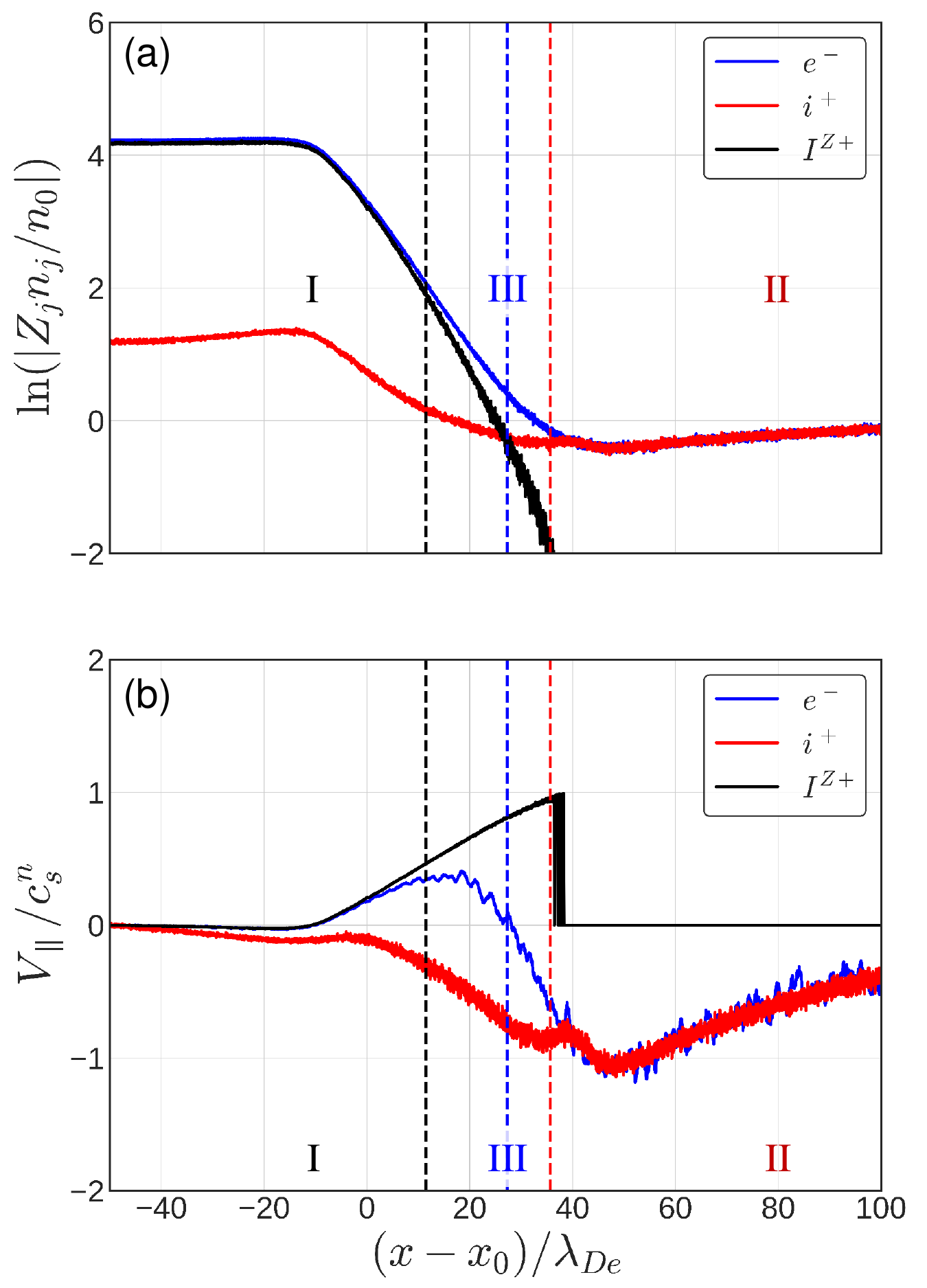}
    \caption{The absolute values of charge density ($Z_j n_j /n_0$
      with $j=I,i,e$) in logarithm~(a) and the parallel flow velocity~(b) at $t\cdot \omega_{pe} = 678$ for the same simulations as
      Fig.~\ref{fig:fig3}. The region to the left of the black dash
      line (where $|n_e-Zn_I|= 0.1~n_e$) is the impurity dominant
      region~(I) where $Zn_I\gg n_i$, the region to the right of red
      dash line (where $|n_e-n_i|= 0.1~n_e$) is the deuterium ion
      dominant region~(II) $Zn_I\ll n_i$, and the interval between
      them is the buffer zone~(III) where the contribution of the two
      ions become comparable $Zn_I\sim n_i$.
    The blue dash line is the impurity front where $Zn_I=n_i$. \textcolor{black}{The parallel flow velocity is normalized by the nominal impurity ion sound speed $c_s^n\equiv \sqrt{ZT_0/m_I}$.}}
    \label{fig:fig4}
\end{figure}

As shown in Fig.~\ref{fig:fig3}, the electron and impurity density in
the early stage propagate steadily in time, and we can define an
impurity front that characterizes how far the impurities can
propagate. Such definition of an impurity front should reflect the
physics that underlies the impurity acceleration, i.e., the ambipolar
electric force dominating over the friction force. Therefore, to
quantify the whereabout of the impurities, we separate the mixing zone
into three regions based on the charge density of the impurity and
deuterium as shown in Fig.~\ref{fig:fig4}: (I) Near the pellet, the
impurity ion dominates $Zn_I\gg n_i$; (II) near the hot surrounding
plasma, the Deuterium dominates $Zn_I\ll n_i$; and (III) between them,
impurity and Deuterium contribute equally $Zn_I\sim n_i$. If we
consider the continuity equations of electrons and impurities
\begin{align}
  \frac{\partial n_{e,I}}{\partial t} + \frac{\partial}{\partial
    x}\left(n_{e,I}V_{e,I\parallel}\right) =
  0,\label{eq:impurity-continuity}
\end{align}
the quasi-neutrality condition, $Zn_I+n_i\approx n_e,$ combined with
ambipolar transport constraint, would determine the parallel plasma
flow as $V_{e\parallel}\approx V_{I\parallel}$, $V_{e\parallel}\approx
(Zn_IV_{I\parallel}+ n_iV_{i\parallel})/(Zn_I+n_i)$, and
$V_{e\parallel}\approx V_{i\parallel}$ in these three regions as shown
Fig.~\ref{fig:fig4}(b).
We hereby define the impurity front (IF) location as where \begin{align}
    Zn_I(\text{IF}) = n_i(\text{IF}). \label{eq:impurity_front_def}
\end{align}

\subsection{Self-similar solution of impurity assimilation\label{sec:self-similar}}

\begin{figure}
    \centering
    \begin{minipage}[b]{0.45\textwidth}
    \includegraphics[width=1\textwidth]{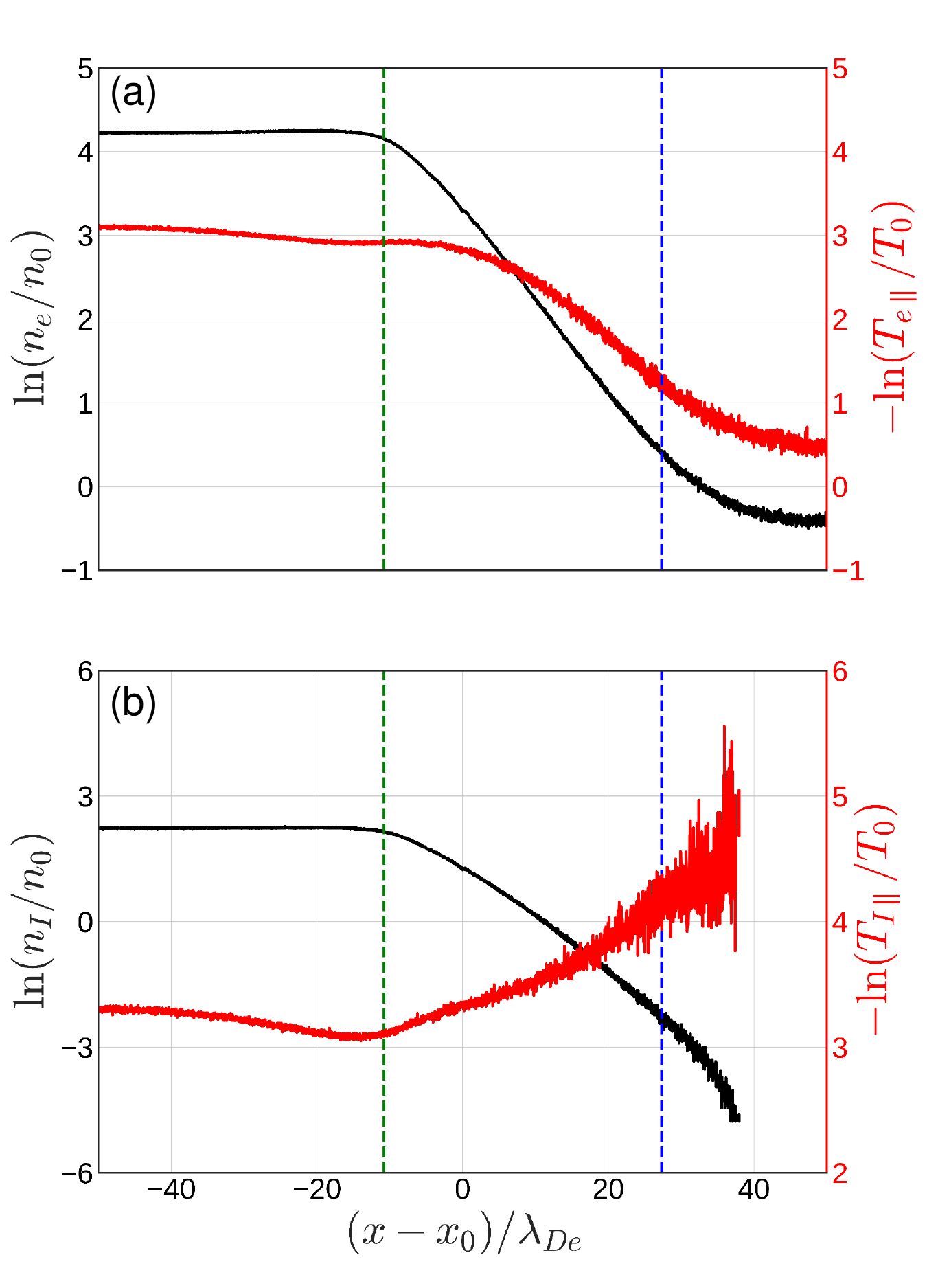}
    \end{minipage}
    \caption{The density and parallel temperature profiles (in
      logarithm) of electrons and impurities at $t\cdot \omega_{pe} =
      678$ from the same simulation as Fig.~\ref{fig:fig3}. The dashed
      blue line is the impurity front and green line is for
      $V_{I\parallel}=0$. From the curves, the fitting parameters,
      between the green and blue dashed lines, in
      Eq.~(\ref{eq:temperature-closure}) are
      \textcolor{black}{$\alpha_e= -0.49$ and $\alpha_I= 0.26 $.}}
    \label{fig:fig5}
\end{figure}

For impurity acceleration in the impurity dominant region (I),
the deuterium ion density is negligible and so is
the ion friction force on the impurities. The impurity
momentum equation of Eq.~(\ref{eq:impurity_momentum_full}),
combined with electron force balance, becomes
\begin{equation}
     m_In_I\frac{dV_{I\parallel}}{dt}+\frac{\partial
       p_{I\parallel}}{\partial x} +\frac{\partial
       p_{e\parallel}}{\partial x} =
     0.\label{eq:impurity_momentum_new}
\end{equation}
This equation combines with the impurity continuity equation of
Eq.~(\ref{eq:impurity-continuity}) and the quasi-neutrality condition
$n_e\approx Zn_I$ to form a complete set of equations if we know the
electron and impurity temperatures, which depend on the heat flux. The
physics of impurity front propagation can be elucidated with a
self-similar solution where all the plasma state variables are
functions of a self-similar variable $U\equiv x/t$. For simplicity, we
assume that the temperatures develop the comparable length scale with
the impurity density as
\begin{equation}
    \frac{\partial \ln T_{e,I\parallel}}{\partial x}
    =(\alpha_e,\alpha_I)\frac{\partial \ln n_{e,I}}{\partial
      x},\label{eq:temperature-closure}
\end{equation}
with $-1<\alpha_{e}<0$ and $\alpha_I>0$ accounting for the electron
heating and impurity decompressional cooling as shown in
Fig.~\ref{fig:fig5}. Notice that $\partial \ln n_I/\partial x=\partial
\ln n_e/\partial x$ under the quasi-neutrality condition with fixed
$Z$.  \textcolor{black}{In the simulations, the variation of
$\alpha_{e,I}$ with time (after the arrival of hot electrons from
  the other side of the pellet) is tiny and $\alpha_e= -0.49$ and
  $\alpha_I= 0.26 $} in Fig.~\ref{fig:fig5}. With these gradient
length scale scalings, we have
\begin{align}
    V_{I\parallel} =U+ c_s,\label{eq:vi-U-relation}
\end{align}
where the impurity ion sound speed is
\begin{align}
    &c_s = \sqrt{\frac{(1+\alpha_I)T_{I\parallel}+ (1+\alpha_e)ZT_{e\parallel}}{m_I}},\label{eq:impurity_ion_sound_speed}
\end{align}
which takes into account the transport physics by including
$\alpha_{e,I}$.
Since $Z\gg 1$ and $T_{e\parallel}\gg
T_{I\parallel}$ due to the electron heating and impurity
decompressional cooling as shown in Fig.~\ref{fig:fig5}, $c_s$ is
dominated by the electron temperature, reinforcing the fact that the
electric force is the main drive for impurity
acceleration/assimilation.

For isothermal plasmas, $c_s$ is the standard local impurity
ion sound speed. Therefore, a natural way to normalize the plasma flow
and impurity front speed would be a nominal impurity ion sound speed
defined using the hot surrounding plasma temperature $T_0,$
\begin{align}
  c_s^n\equiv \sqrt{ZT_0/m_I}. \label{eq:impurity-assimilation-nominal-speed}
\end{align}
As we shall see shortly from the simulation data, $c_s^n$ turns out to
be the characteristic speed of impurity front, with
$U_s=U(\text{IF})\sim 0.5 c_s^n.$ This provides a quick quantitative estimate
on how fast impurity ion assimilation would occur toroidally in a
tokamak or stellerator.

With Eq.~(\ref{eq:impurity_ion_sound_speed}), the impurity continuity
equation of Eq.~(\ref{eq:impurity-continuity}) becomes
\begin{align}
    \frac{\partial V_{I\parallel}}{\partial x}= -c_s\frac{\partial \ln n_I}{\partial x}.\label{eq:impurity_continuity-new-new}
\end{align}
Integrating it from the location where the flow velocity $V_{I\parallel}
= 0$ ($n_I\approx n_{I0}$), to the impurity front, we obtain
\begin{align}
    & V_{I\parallel}(\text{IF})=\eta\bar{c}_s
,\label{eq:self_similar_flow_velocity}
\end{align}
with
\begin{equation}
   \eta \equiv \ln\left(\frac{n_{I0}}{n_I(\text{IF})}\right). \label{eq:eta} 
\end{equation}
To obtain Eq.~(\ref{eq:self_similar_flow_velocity}), we have employed
$\partial \ln n_{I}/\partial x =const. $ as shown in
Fig.~\ref{fig:fig5}, and thus $\bar{c}_s\equiv\langle c_s\rangle_x$ is
the spatial average of the impurity ion sound speed from
$V_{I\parallel}=0$ to the impurity front. Notice that the plasma
density appears only through its logarithm in $\eta$ and $\eta \approx
\ln (n_{e0}/n_0)\approx \ln (T_0/T_{cold})$ for $T_0\gg T_{cold}$.

Substituting Eq.~(\ref{eq:self_similar_flow_velocity}) into
Eq.~(\ref{eq:vi-U-relation}), we obtain the impurity front speed
\begin{align}
    &U(\text{IF}) = \eta\bar{c}_s-c_s(\text{IF}).\label{eq:impurity_front_velocity}
\end{align}
It shows that the impurity front speed is proportional to the
impurity's charge mass ratio $Z/m_I$, the largest value of which for
fully ionized impurities is the same as that of the Deuterium
$1/m_i$. Therefore, the impurity front would stay behind the cooling front.

\begin{figure}
    \centering
    \begin{minipage}[b]{0.45\textwidth}
    \includegraphics[width=1\textwidth]{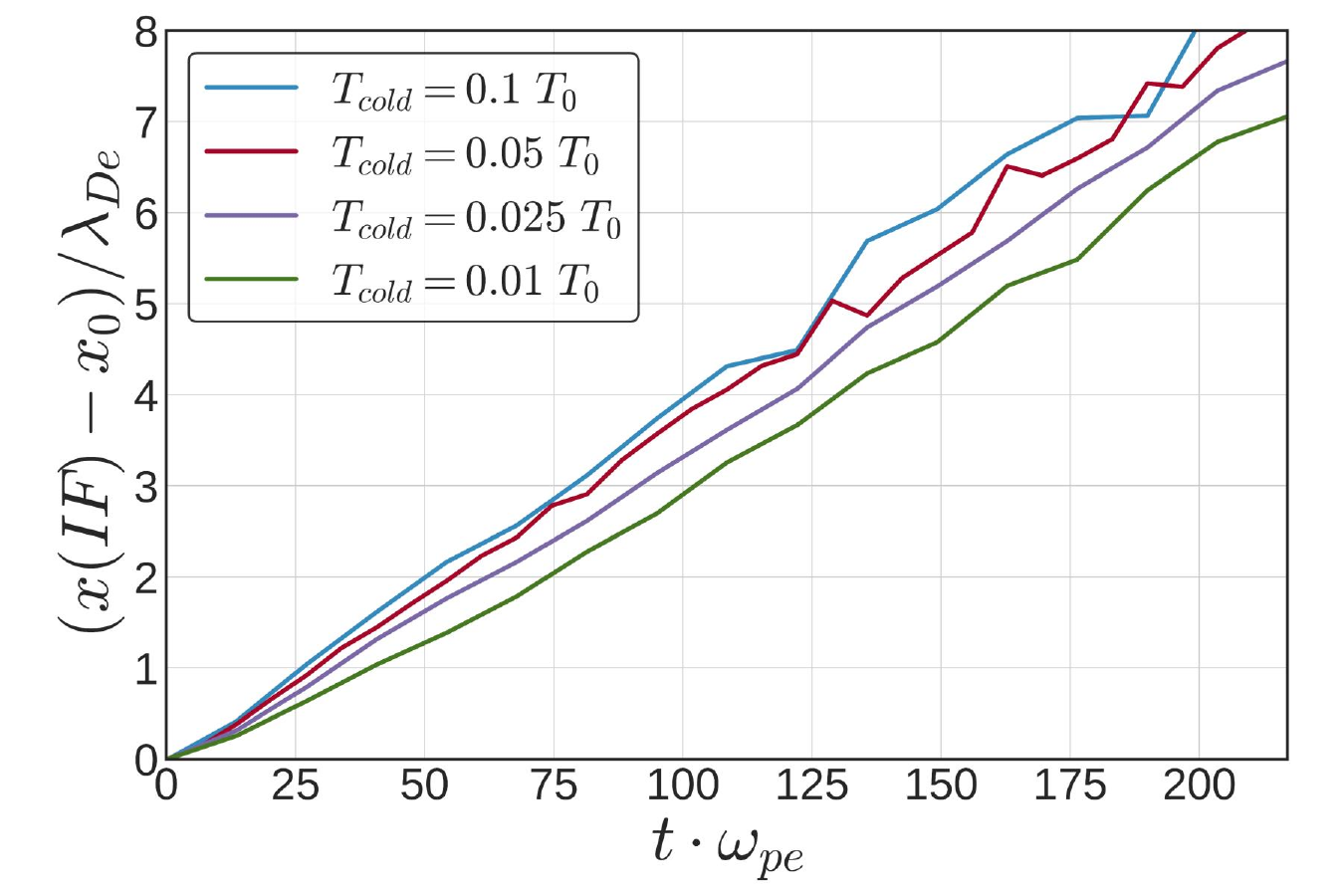}
    \end{minipage}
    \caption{Numerical impurity front location for different pellet temperature $T_{cold}$ with $Z=7$. }
    \label{fig:fig6}
\end{figure}

Equation~(\ref{eq:impurity_front_velocity}) also reveals that the
impurity front speed is actually dominated by the electron
temperature, and hence the key problem is how the electron temperature
is developed at and behind the impurity front, which will be discussed
in next section. Before answering this question, we first show the
numerical verification of the self-similar solution in
Fig.~\ref{fig:fig6} and Table~\ref{table:parameters}, where
\textcolor{black}{the relative error between the impurity front speed
  from the simulations and the self-similar solutions is within
  $25\%$}. Figure~\ref{fig:fig6} also reaffirms that the impurity front
propagates nearly steadily. But most importantly, Fig.~\ref{fig:fig6}
and Table~\ref{table:parameters} show that the impurity fronts have a
weak dependence on the cold pellet temperature, where the largest
difference of the impurity front speed (between $T_{cold}=0.1~T_0$ and
$T_{cold}=0.01~T_0$ cases) \textcolor{black}{is only near $ 25\%$}
regardless of the ion transparentness, which highlights again the
importance of the electron temperature at the impurity front ($c_s(\textrm{IF})$), and
behind the impurity front ($\bar{c}_s$). Notice that the electron
density enters only through its logarithm that has a dependence on the
pellet temperature via $\eta\approx \ln(T_{0}/T_{cold})$. Such
temperature dependence in $\eta$ will actually be canceled by
$\bar{c}_s$ if we consider Eq.~(\ref{eq:temperature-closure}) and
assume linear profiles of $\ln(n_e)$ and $\ln(T_{e\parallel})$, for which one
obtains $\bar{c}_s\propto \langle \sqrt{T_{e\parallel}}\rangle_x \propto
\sqrt{T_{e\parallel}(\textup{IF})}[1-\exp(\alpha_e\eta/2)]/(\alpha_e\eta)
$. Notice that from Table \ref{table:parameters}, $\alpha_e$ and
$T_{e\parallel}(\textup{IF})\propto c_s(\textup{IF})^2$ only slightly vary
 with pellet plasma temperature, so the pellet temperature enters only through a small factor
of $\exp(\alpha_e\eta/2)$.

\begin{table}
      \centering
        \begin{tabular}{|c|c|c|c|c|c|c|c|}
        \hline
        $T_{cold}/T_0$& $U_{s}/c_s^n$ &  $U_{s}/U_{t}$ & $\eta$ &$\bar{c}_s/c_s^n$ &$c_s(\text{IF})/c_s^n$ & $\alpha_e$ & $\alpha_I$ \\
        \hline
        0.1     & 0.60   & 1.18  & 2.9   & 0.33 & 0.45 & -0.48 & 0.43 \\
        \hline
        0.05    & 0.57  & 1.07  & 3.7  & 0.25 & 0.39 & -0.49 & 0.36  \\
        \hline
        0.025   & 0.53  & 0.92  & 4.5  & 0.21 & 0.37 & -0.52 & 0.31 \\
        \hline
        0.01    & 0.48  & 0.78   & 5.4  & 0.18 & 0.36 & -0.50 & 0.22 \\
        \hline
        \end{tabular}
    \caption{ Simulation results for different pellet temperature $T_{cold}$ of the impurity front speed $U_s=U(\textup{IF})$ and its normalization by the theoretic value $U_t=\eta\bar{c}_s-c_s(\text{IF})$ in Eq.~(\ref{eq:impurity_front_velocity}). The parameters are taken at $t\cdot\omega_{pe} = 217$.
    }
    \label{table:parameters}
\end{table}

\section{Electron temperature in determining the impurity front speed \label{sec:Te}}

This section investigates the electron temperature that determines the
impurity front speed through $c_s(\textrm{IF})$ and $\bar{c}_s$. As we
see from Table \ref{table:parameters}, both $c_s(\textrm{IF})$ and
$\bar{c}_s$ have a weak dependence on the pellet plasma temperature
and hence so does the impurity front speed, regardless of the hot ion
transparency in the pellet plasma. In other words, the impurity front
speed is mainly governed by the surrounding hot plasma
temperature. This interesting outcome is further highlighted by
comparing the results with the collisionless simulation in Table
\ref{table:parameters-0.01}, where the collisionless simulation yields
nearly the same impurity front speed as the collisional one with a
deviation of $\lesssim 25\%$. Notice that the set-up of the
collisionality mainly affect the plasma behind the cooling front
rather than the surrounding hot plasma.

\begin{table}
      \centering
        \begin{tabular}{|c|c|c|c|c|c|}
        \hline
        & $U_s/c_s^n$& $U_s/U_t$ & $\eta$ &$\bar{c}_s/c_s^n$ &$c_s(\text{IF})/c_s^n$  \\
        \hline
        collisional         & 0.48   & 0.77   & 5.4  & 0.18  & 0.35   \\
        \hline
        collisionless       & 0.35   & 0.94   & 4.7  & 0.13  & 0.24  \\
        \hline
        \end{tabular}
    \caption{ Collisional and collisionless simulation results for $T_{cold}=0.01~T_0$ at $t\cdot\omega_{pe} = 217$, 
    where the collisional simulation is the same as the one in Table~\ref{table:parameters}.
    }
    \label{table:parameters-0.01}
\end{table}

\begin{figure}
    \centering
    \begin{minipage}[b]{0.45\textwidth}
    \includegraphics[width=1\textwidth]{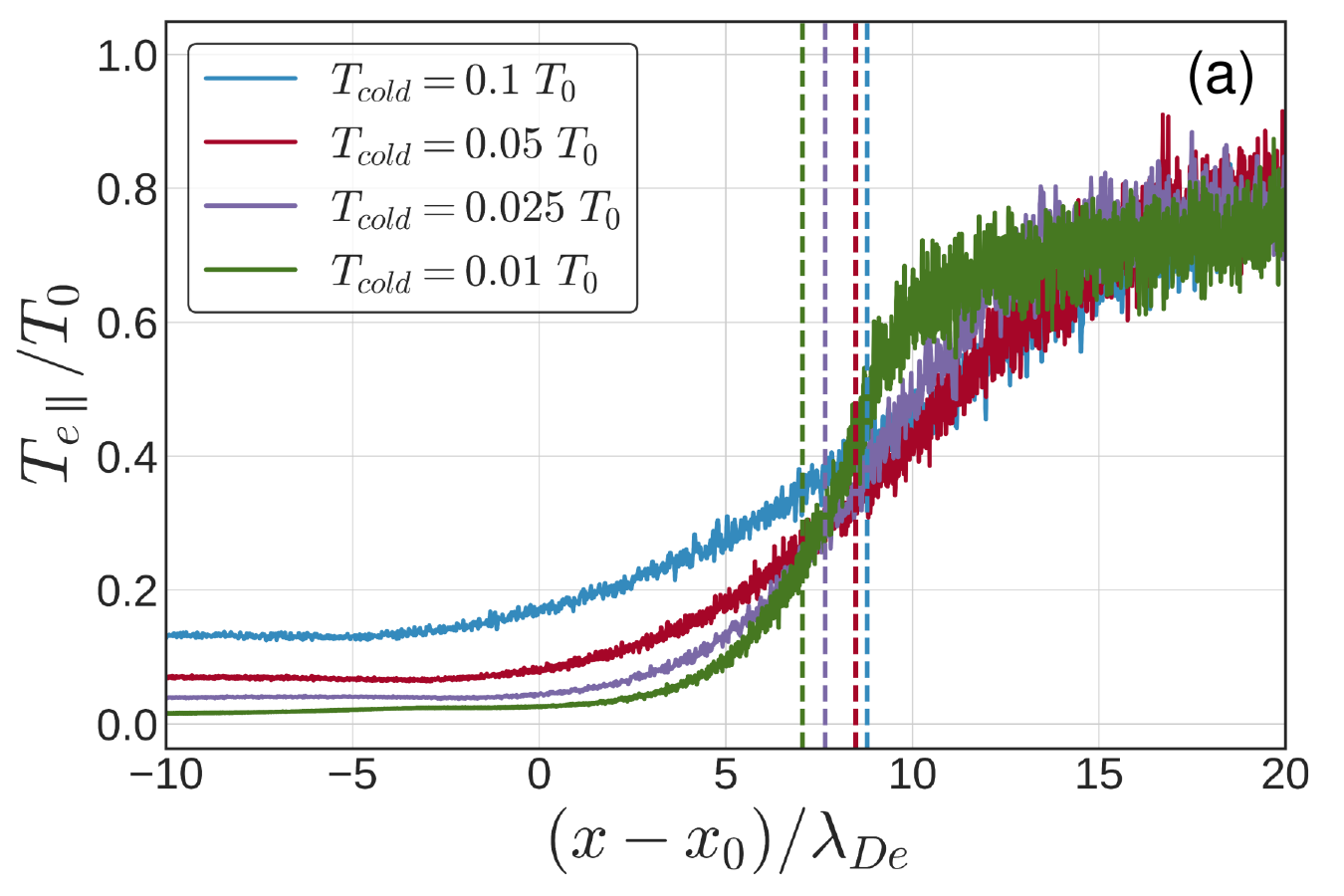}
    \end{minipage}
    \begin{minipage}[b]{0.45\textwidth}
    \includegraphics[width=1\textwidth]{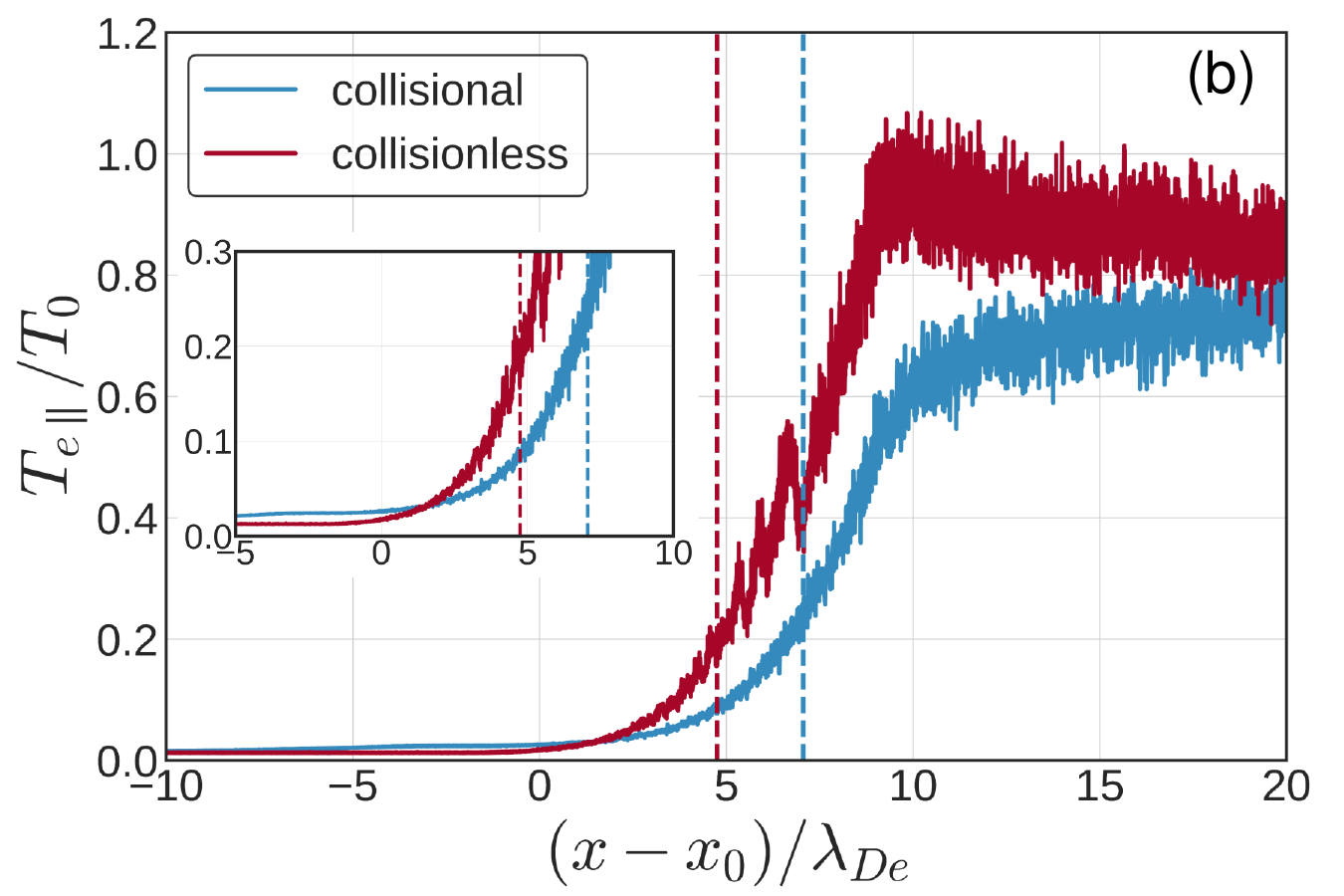}
    \end{minipage}
    \caption{Electron parallel temperature for different cold pellet temperature with collisions (a) and for different collisionality (and their zoom-in) with $T_{cold}=0.01~T_0$ (b). The dash lines in corresponding colors denote the impurity front.  These snapshots are taken at $t\cdot\omega_{pe} = 217$.}
    \label{fig:fig7}
\end{figure}

\begin{figure}
    \centering

    \begin{minipage}[b]{0.45\textwidth}
    \includegraphics[width=1\textwidth]{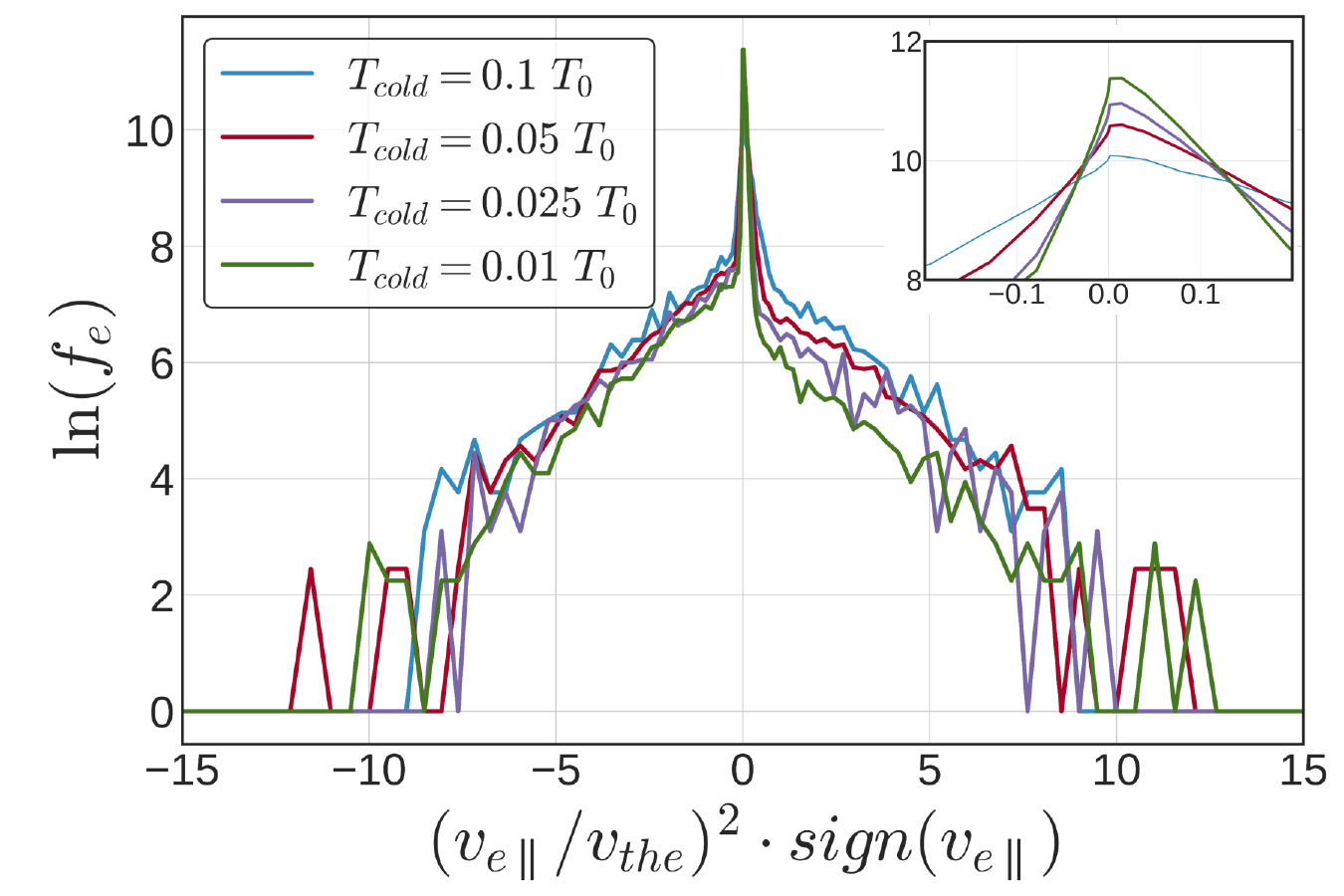}
    \end{minipage}
    \caption{Electron distribution $f_e(v_{e\parallel})$ and their zoom-in at the cold bulk at $2~\lambda_{De}$ behind the corresponding impurity front corresponding to Fig.~\ref{fig:fig7}(a). The hot tail electrons with positive $v_{e\parallel}$ are from the other side of the pellet.}
    \label{fig:fig8}
\end{figure}

It is worth noting that the agreement of the impurity front speed with
the collisionless simulation indicates that the speed of impurity
assimilation is fundamentally a collisionless process that is
dominated by kinetic physics.  This is because near the impurity
front with small impurity density $ Zn_I\sim n_i\sim n_0$, the
upstream hot electrons can nearly-collisionlessly penetrate through
the impurity front and dominate the electron temperature through the
hot electron component.  This can be seen in the electron temperature
profiles in Fig.~\ref{fig:fig7}, in which $T_{e\parallel}$ (and thus $c_s(\textup{IF})$) is nearly
the same at the impurity front for all the cases since the hot tail
electrons will dominate $T_{e\parallel}$ as seen from
Fig.~\ref{fig:fig8}. Notice that Fig.~\ref{fig:fig8} manifests that
the mixing of hot tail and cold bulk electrons is far from a
collisional equilibrium, but with distinctly a hot tail component and
a cold bulk. Therefore, the kinetic physics is of great importance in the impurity assimilation~\cite{arnold2023electron,arnold2023parallel}, which cannot be treated by fluid models.

 On the other hand, $\bar{c}_s\propto \langle
\sqrt{T_{e\parallel}}\rangle_x$ will be dominated by the domain near
the impurity front since the electron temperature has a sharp decrease
behind the impurity front due to the increase of the fraction of the
cold pellet electrons (e.g., see Fig.~\ref{fig:fig7}). This mechanism
can be quantified if we consider again
Eq.~(\ref{eq:temperature-closure}) and assume linear profiles of
$\ln(n_e)$ and $\ln(T_{e\parallel})$, where $25\%$ of the domain near
the impurity front contributes $50\%$ of the integral for
$\bar{c}_s$. Therefore, both $\bar{c}_s$ and $c_s(\textup{IF})$ are
dominated by the surrounding hot plasmas and so is the impurity front
speed as seen from Eq.~(\ref{eq:impurity_front_velocity}).

\begin{figure}
    \centering
    \begin{minipage}[b]{0.45\textwidth}
    \includegraphics[width=1\textwidth]{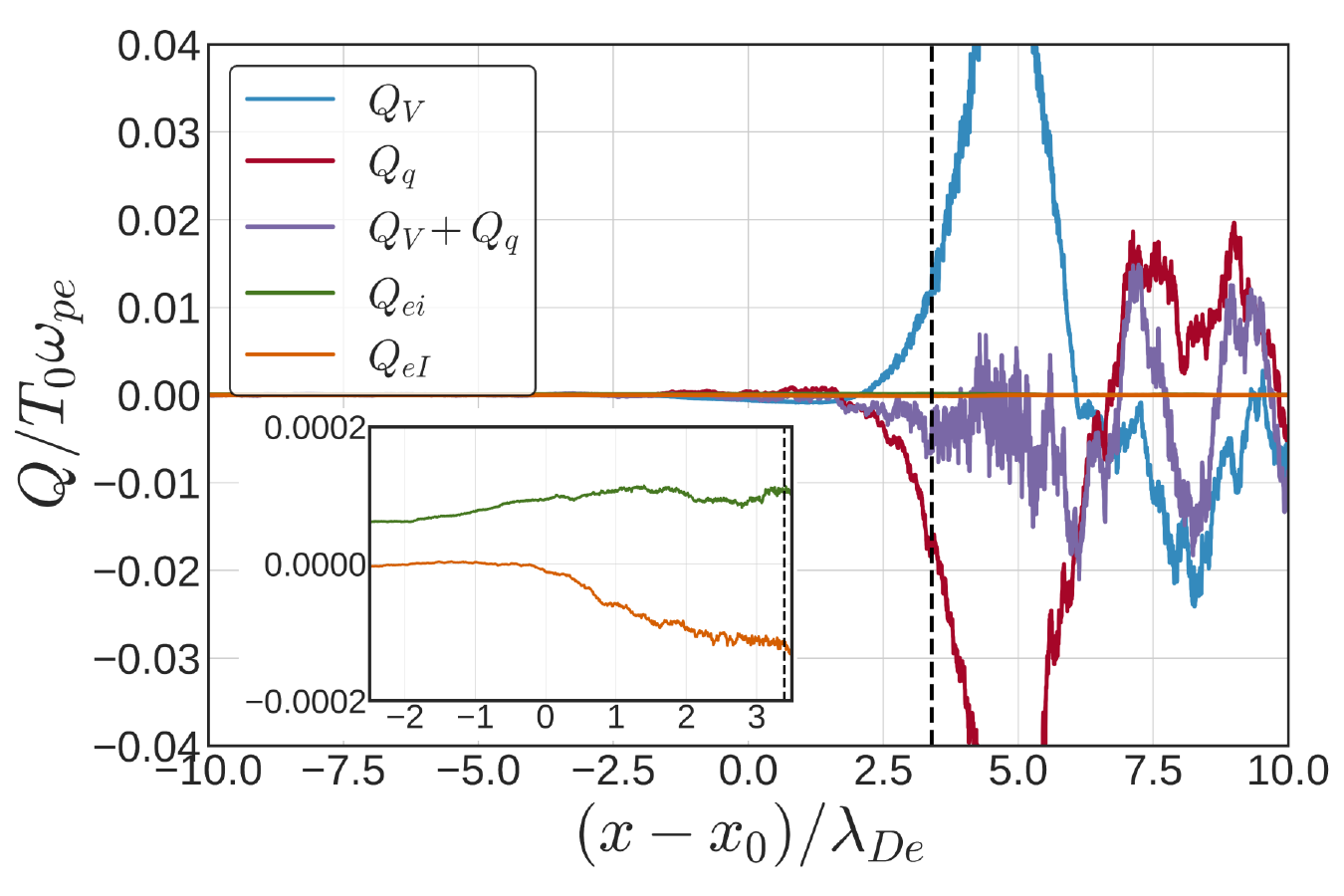}
    \end{minipage}
    \caption{Heat fluxes that determine the electron temperature evolution in Eq.~(\ref{eq-electron-energy}) for $T_{cold}= 0.025~T_0$ at $t\cdot\omega_{pe} = 101$. The zoom-in shows the collisional heat fluxes, which is two orders of magnitude smaller than the convective and conductive heat fluxes.}
  \label{fig:fig9}
\end{figure}

The collisionless nature of the electron temperature near the impurity
front can also be illustrated by considering the energy fluxes that
determine the temperature evolution
\begin{align}
  \frac{\partial  T_{e\parallel}  }{\partial t}=Q_V+Q_q+Q_{ei}+Q_{eI},\label{eq-electron-energy}
\end{align}
where $Q_V=-V_{e\parallel}\partial T_{e\parallel}/\partial x -
2T_{e\parallel}\partial V_{e\parallel} /\partial x$ reflects the
convective energy flux, $Q_q=- (\partial q_{en}/\partial x)/n_e$
denotes the conductive energy flux, and $Q_{ei,I}$ are due to the
collisions with the deuterium and impurity. Notice that $Q_V$ and
$Q_q$ represents the collisionless heat fluxes. As shown in
Fig.~\ref{fig:fig9}, in the domain just behind the impurity front, the
convective and conductive fluxes, despite their opposite contributions, dominate over the collisional fluxes. From the collisional fluxes,
which is two orders of magnitude smaller than the convective and
conductive heat fluxes, the electrons will gain energy from the
deuterium but lose energy to the impurity. In physical terms, the main
energy exchange channel is that the cold electrons gain energy from
the hot ions while the hot electrons lose energy to the
impurities. Another observation is that the electron heating by ions
will dominate over the cooling by impurities in the region where the
collisionless fluxes are small, which will speed up the impurity
acceleration by enhancing $\bar{c}_s$ (e.g., see
Fig.~\ref{fig:fig7}(b) for the comparison of the collisionless and
collisional temperature).

\section{\label{sec:summary}Conclusions}

In conclusion, we have investigated the impurity assimilation into the
surrounding hot plasma along the magnetic field line, which would
determine the uniformity of the high-Z impurities and hence the
radiation peaking factor around the torus. 1D3V first-principles kinetic simulations
show that the impurities propagate steadily
into the surrounding hot plasma with the alignment of the impurities
of different charge states due to the strong collisionality of the
pellet ions. This suggests that to leading order, the assimilation of impurities can be
described by an averaged charge state, highlighting the fact that the
parallel transport of impurities is dominated by the ambipolar electric force
through the electron pressure gradient.

Based on these observations, an impurity front has been defined as
where the charge density of the impurities equals to that of the
plasma ions. With the help of a self-similar solution, we find that
the impurity front is behind the plasma cooling front due to the
smaller charge-mass-ratio of the impurity ions, and thus the
ambipolarity-regulated plasma power flux from the hot plasma plays a
critical role in the impurity assimilation process. We have shown that
the impurity front speed $U_s$ has a weak dependence on the pellet
temperature and system collisionality (for single charge state
impurity), and the impurity assimilation is a collisionless process
that is dominated by the surrounding hot plasma. Specifically, $U_s$
is at the order of a nominal impurity ion sound speed $U_s\sim c_s^n,$
with $c_s^n$ defined in
Eq.~(\ref{eq:impurity-assimilation-nominal-speed}) using the
temperature $T_0$ of the surroudning hot plasma, the averaged charge
state of the impurity ions, and the impurity ion mass. Such
collisionless nature has been illustrated by the fact that the
electron temperature near the impurty front is primarily determined by
the hot tails and collisionless heat fluxes.

\begin{acknowledgments}
We thank the U.S. Department of Energy Office of Fusion Energy Sciences and Office of Advanced Scientific Computing Research for support under the Tokamak Disruption Simulation (TDS) Scientific Discovery through Advanced Computing (SciDAC) project, and the Base Theory Program, both at Los Alamos National Laboratory (LANL) under contract No. 89233218CNA000001. H.M. is funded by the UC-National Laboratory In-Residence Fellowship under the reward number L22GF4528. This research used
resources of the National Energy Research Scientific Computing Center (NERSC), a U.S. Department of Energy Office of Science User Facility operated under Contract No. DE-AC02-05CH11231 and the Los Alamos
National Laboratory Institutional Computing Program, which is supported by the U.S. Department of Energy National Nuclear Security
Administration under Contract No. 89233218CNA000001. Special thanks to Dr. Jun Li for the valuable suggestions on the simulations.
\end{acknowledgments}


\bibliography{aipsamp}

\end{document}